\documentclass[twocolumn]{article}
\usepackage[utf8]{inputenc}

\usepackage{graphicx}
\usepackage{color,soul}
\usepackage{braket}
\usepackage{amsmath}
\usepackage{cleveref}
\usepackage{amssymb}
\usepackage{biblatex}
\usepackage{caption}
\usepackage{authblk}

\title{\vspace{-3.5cm}\textbf{GA4QCO: Genetic Algorithm for Quantum Circuit Optimization}}

\author[1]{\small \textbf{Leo Sünkel} \thanks{leo.suenkel@ifi.lmu.de}}
\author[2]{\textbf{Darya Martyniuk} \thanks{darya.martyniuk@fokus.fraunhofer.de}} 
\author[2]{\textbf{Denny Mattern}\thanks{denny.mattern@fokus-extern.fraunhofer.de}}
\author[2, 3]{\textbf{Johannes Jung} \thanks{johannes.jung@fokus.fraunhofer.de}}
\author[2, 3]{\textbf{Adrian Paschke}}

\affil[1]{\small Institute for Informatics, LMU Munich, Germany}
\affil[2]{Data Analytics Center, Fraunhofer FOKUS, Berlin, Germany}
\affil[3]{Institute for Computer Science, Freie Universität Berlin, Germany}

\date{}

\begin{document}

\maketitle
\begin{abstract}
 The design of quantum circuits is often still done manually, for instance by following certain patterns or rule of thumb. While this approach may work well for some problems, it can be a tedious task and present quite the challenge in other situations. Designing the architecture of a circuit for a simple classification problem may be relatively straightforward task, however, creating circuits for more complex problems or that are resilient to certain known problems (e.g. barren plateaus, trainability, etc.) is a different issue. Moreover, efficient state preparation or circuits with low depth are important for virtually most algorithms. In attempts to automate the process of designing circuits, different approaches have been suggested over the years, including genetic algorithms and reinforcement learning. We propose our GA4QCO framework that applies a genetic algorithm to automatically search for quantum circuits that exhibit user-defined properties. With our framework, the user specifies through a fitness function what type of circuit should be created, for instance circuits that prepare a specific target state while keeping depth at a minimum and maximizing fidelity. Our framework is designed in such a way that the user can easily integrate a custom designed fitness function. In this paper, we introduce our framework and run experiments to show the validity of the approach.
\end{abstract}

\thanks{\footnotesize This work has been submitted to the IEEE for possible publication. Copyright may be transferred without notice, after which this version may no longer be accessible.}

\section{Introduction}
While quantum computing has been gaining momentum in recent years, the capabilities of quantum computers of the current so-called Noisy-Intermediate-Scale-Quantum (NISQ) \cite{preskill2018quantum} era are still severely limited. 
That is, current quantum computers contain relatively few qubits that are error-prone, circuits are restricted to a low depth and algorithms are run without any error-correction. 
To overcome these limitations, hybrid algorithms, i.e., algorithms that partially run on classical and quantum computers, are often used instead of purely quantum algorithms~\cite{mitarai2018quantum,schuld2020circuit}.
Although quantum supremacy was claimed in 2019~\cite{arute2019quantum}, it is yet to be seen whether a practical advantage can be produced in the NISQ-era.

Putting aside whether NISQ-algorithms will produce the highly anticipated quantum advantage, other fundamental questions must also be addressed. For instance, how does one construct or design a quantum circuit for a specific task? What constitutes a good circuit? Why choose one ansatz over another? Quantum circuits are often hand-crafted, i.e., one manually comes up with a design of a particular circuit. This can for example be done by following standard practices. However, designing efficient and powerful circuits for NISQ-devices is a non-trivial task. Naturally the question arises whether this process can be automated, that is, could an algorithm design novel quantum circuits for a given task? This question, as we will see, is not new and different approaches in this regard have been investigated by the research community. 

In this paper, we introduce our GA4QCO framework, a framework that uses a genetic algorithm to design completely new circuits and to optimize existing ones. More importantly, our proposed framework allows its users to define a custom fitness function to automatically search for quantum circuits matching a user-defined criteria. We start by briefly discussing related work and different approaches that have been proposed for the task of searching for quantum circuits by the research community. In Section \ref{sec:background} we give a short recap of the fundamentals of genetic algorithms and quantum computing required to understand the core components of our proposed method. We introduce our framework in Section \ref{sec:our_framework}. We apply our framework to different settings relevant in the NISQ-era and discuss our experimental setup and results in Section \ref{sec:experiments}. We give an outlook on future work in Section \ref{sec:outlook} and a conclusion in Section \ref{sec:conclusion}.

\section{Related Work} \label{sec:related_work}
The idea of automatically searching and designing quantum circuits has been around for a while, and a prominent approach in this area is to apply genetic algorithms. Williams and Gray use a genetic programming algorithm to design quantum circuits in \cite{williams1998automated}. In their approach, the algorithm is given a target unitary matrix and the goal is to find circuits whose unitary matrix is minimally different than the target. They apply their algorithm to the problem of creating circuits for quantum teleportation. Rubinstein proposes a genetic programming approach for evolving circuits that result in maximally entangled qubits in \cite{rubinstein2001evolving}. Lukac and Perkowski propose a genetic algorithm with the goal to evolve circuits that correspond to a specified target unitary matrix \cite{lukac2002evolving}. More recently, Ding and Spector proposed a genetic algorithm for quantum architecture search for parameterized quantum circuits in the domain of reinforcement learning \cite{ding_evolutionary_2022} and Rattew et al. propose their evolutionary variational eigensolver in \cite{rattew2019domain}. 
Zhang and Zhao propose an evolutionary algorithm for the task of circuit architecture search and apply it to classification \cite{zhang2022evolutionary}.

Other algorithms and methods have also been proposed for this task. Du et al. introduced their quantum architecture search (QAS) algorithm in \cite{du_quantum_2022}, which is an algorithm to automatically design quantum circuits with improved learning behavior for variational quantum algorithms. 
In \cite{lu_markovian_2021} Lu et al. propose their Markovian quantum neuralevolution (MQNE) algorithm, which searches for quantum circuits in the context of quantum machine learning. 
Pirhooshyaran and Terlaky investigate the use of three different approaches, i.e., random search, reinforcement learning and Bayesian optimization to search for quantum circuits. In their approach, the trainability of the circuits is taken into account during the search \cite{pirhooshyaran_quantum_2021}. 
Wu et al. introduced a compilation and circuit optimization framework named QGo in \cite{wu_qgo_2022}. QGo can produce an optimized circuit for a target device. Chivilikhin et al. \cite{chivilikhin_mog-vqe_2020} propose a technique that, among other things, optimizes the topology of a quantum circuit for a hardware-efficient variational quantum eigensolver through the use of a multi-objective genetic algorithm. The genetic algorithm aims to minimize the number of CNOT gates in a circuit as well as the energy of the system. Kuo et al. employ deep reinforcement learning to find quantum circuits that generate a specified target state from an initial state~\cite{kuo_quantum_2021}. 

\section{Background} \label{sec:background}
In this section, we will briefly recap the fundamentals of genetic algorithms (GAs) and quantum computing (QC) that are required to understand the core concepts of our proposed framework, which we will introduce in Section \ref{sec:our_framework}. For more comprehensive introductions to GAs we refer to \cite{eiben_evolutionary_2005} and to QC to \cite{nielsen2010quantum}.
\subsection{Genetic Algorithms}
GAs are meta-heuristic optimization algorithms that incorporate many methods and concepts inspired by biological Darwinian evolution. That is, they are algorithms inspired by nature to tackle a wide range of problems that may appear in a plethora of domains.
A GA is made of a number of building blocks and we will introduce the essential ones next.

\textbf{Individual:} A single solution is represented by an individual. An individual may consist of several properties, however, we can think of it as an representation or encoding of a single solution for the problem at hand.

\textbf{Population:} Every GA consists of a population, which contains a number of individuals. It can be thought of as a list of individuals.

\textbf{Crossover:} Throughout the evolutionary process new individuals (i.e. solutions) are created through the use of crossover operations. While there exist a variety of different crossover operations, their common purpose is to create a new individual, often referred to as the child, by combining the properties of two existing individuals, often referred to as the parents. This recombination does not always produce better individuals, however, through the evolutionary process, which is inspired by the idea of survival of the fittest, bad solutions and traits should be neglected and better solutions should ultimately be preferred and thrive.

\textbf{Mutation:} Random mutations are applied to individuals resulting in a slightly different solution. A mutation could for example be a simple bit-flip, however, more elaborate mutations may be applied.

\textbf{Generation:} A GA runs for a number of generations. In each generation, new solutions are created via crossover and mutation. At the end of each generation, some individuals are discarded while the rest survive and take part in the next generation.

\textbf{Selection:} Selection pressure is applied in two occasions.
For one, there is the parent selection, i.e., the process of determining what individuals are allowed to take part in the crossover operations in a particular generation. Secondly, there is the survivor selection, i.e., the process of determining which individuals are going to be part of the next generation and which are discarded.

\textbf{Fitness:} Each individual has a fitness value, which determines how good a particular solution is. It drives the evolutionary process and allows the ranking of individuals.

\textbf{Evolutionary process:} A GA starts with a population of random individuals and then runs for a number of generations with the goal to find individuals with the highest fitness value. In each generation, new individuals (children) are created by applying crossover operations on existing individuals (parents). Mutations are applied randomly to newly created solutions. Individuals are ranked by fitness determined by a problem specific fitness function. At the end of each generation, weaker individuals are replaced by children thus creating the population for the next generation. This process repeats until the specified number of generations is reached. As randomness is part of the algorithm, convergence is not guaranteed, and each run may return different solutions. \cite{eiben_evolutionary_2005,eiben2015introduction}

These are the basic building blocks of a GA and we will see how we construct our algorithm for the search for quantum circuits in Section~\ref{sec:our_framework}. However, we will first introduce the basics of QC.

\subsection{Quantum Computing}

\begin{figure*}[t!]
\begin{center}
    \includegraphics[width=1\textwidth]{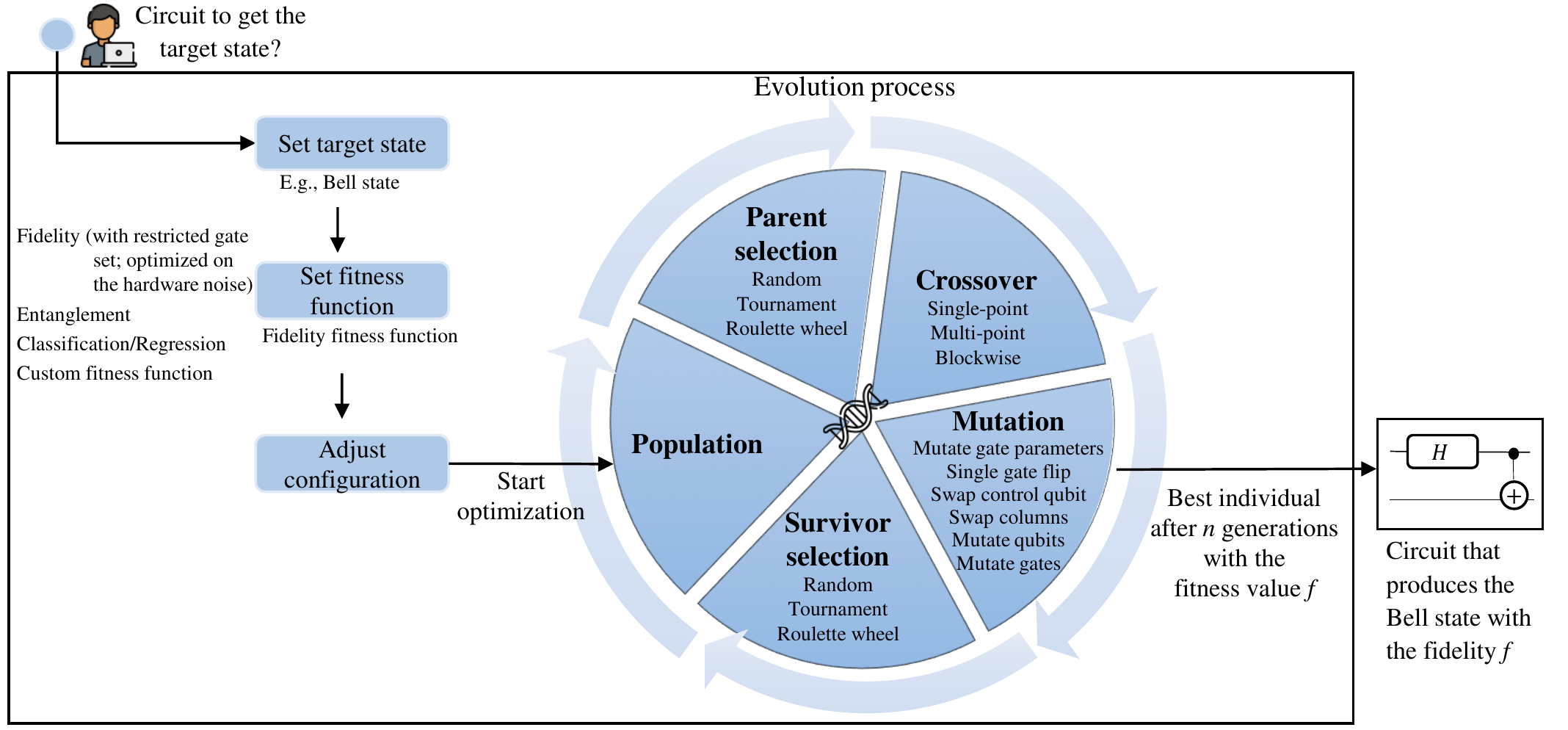}
    \caption {Optimization of quantum circuits in the GA4QCO framework.}
    \label{fig:EvolutionProcess}
    \end{center}
\end{figure*}
Quantum computing is a concept in quantum information theory that utilizes quantum properties to gain an advantage over classical computing for some problems.
Instead of working with classical bits, that can be either zero or one, quantum computers work with so called \textit{qubits}. 
Qubits can be in a superposition of zero and one, which we will denote as~$\ket{0}$~ and~$\ket{1}$.
A qubit state $\ket{\phi}$ can be seen as an unit vector in a two dimensional complex Hilbert space with basis states $\ket{0}= \begin{pmatrix} 1 \\ 0 \end{pmatrix}$ and $\ket{1}=\begin{pmatrix} 0 \\ 1 \end{pmatrix} $ (which we will call the \textit{computational basis}), such that

\begin{align} \label{qubit}
    \ket{\phi} = \alpha_0 \ket{0} + \alpha_1 \ket{1}
    = \alpha_0 \begin{pmatrix} 1 \\ 0 \end{pmatrix} + \alpha_1 \begin{pmatrix} 0 \\ 1 \end{pmatrix} 
\end{align}
where $\alpha_0, \alpha_1 \in \mathbb{C}$ and correspond to the probabilities of collapsing to the states $\ket{0}$ or $\ket{1}$  respectively. Note that $|\alpha_0|^2 + |\alpha_1|^2 = 1$ holds.

A system of multiple qubits can be described as a product space of the single qubit spaces.
For example, if we have two qubits, a general state $\ket{\varphi} \in \mathcal{H} \times \mathcal{H}$ can be described as $\ket{\varphi} = \alpha_{00} \ket{00} + \alpha_{01}\ket{01} +\alpha_{10}\ket{10} +\alpha_{11}\ket{11} $, where now $|\alpha_{00}|^2$ is the probability that the first and second qubit is measured in the $\ket{0}$ state, $|\alpha_{01}|^2$ is the probability that the first state is measured in the $\ket{1}$ state and the second qubit in the $\ket{0}$ state and so on.

A quantum computer can act via \textit{gates} to map an initial quantum state to another state.
Since quantum states are unit vectors in the $2^n$-dimensional Hilbert space~$\mathcal{H}$, the gates must ensure that the resulting state is always a unit vector. That is the reason why gates of a quantum computer are always represented by unitary matrices.
A matrix is unitary iff $U^{\dagger} U = UU^{\dagger} = I$, where $I$ is the identity matrix. $U$ has the property that it maps a unit vector $\ket{\varphi} \in \mathcal{H}$ to another unit vector $U \ket{\varphi} \in \mathcal{H}$.
In theory, any unitary matrix can be applied to the system, but in practice only local gates (gates that act on a subsystem of only a few qubits) are implemented on a quantum computer. This is due to the physical limitations (e.g. noise).

\textit{Entanglement} is another important aspect in QC. Loosely speaking, two qubits are entangled if their states depend on each other. For example, the bell state $\ket{\varphi} = \frac{1}{\sqrt{2}} \ket{00} + \frac{1}{\sqrt{2}} \ket{11}$ is an entangled state: If the first qubit is measured in the $\ket{0}$ state the second qubit also have to be in the $\ket{0}$ state and the same for the $\ket{1}$ state. 
Formally, a two-qubit state $\ket{\varphi} \in \mathcal{H} \times \mathcal{H}$ is in an entangled state if there are no $\varphi_1, \varphi_2 \in \mathcal{H}$ such that $\ket{\varphi} = \ket{\varphi_1} \otimes \ket{\varphi_2}$.

The existence of entangled states on a quantum computer is a fundamental difference to classical computers and a main reason why quantum computing can yield an advantage.
Such an advantage have be proven for various quantum algorithms theoretically. 
For example, the Grover algorithm \cite{grover} for finding elements in an unsorted database is quadratically faster than any classical algorithm and Shors algorithm \cite{Shor_1997} for prime factorization leads even to an exponential speed up.

\section{The GA4QCO-Framework}\label{sec:our_framework}
Our GA based framework optimizes quantum circuits while enforcing specific properties with respect to given constraints, e.g. searches for circuits with minimal depth while still achieving high fidelity. The framework is designed as a toolbox, allowing different components to be combined. Furthermore, custom, i.e., user-defined fitness functions are easily integrated, allowing users to apply the framework to a wide range of problems and domains.

The framework simulates the evolution process similar to the illustration in~\Cref{fig:EvolutionProcess}. The user can control the main properties of the evolution via a property file, e.g. number of individuals in the population, number of generations, probability of the mutation and crossover operations, the depth and width of the circuits, type of the fitness function and others. 

\subsection{Components}\label{sec:components_of_the_framework}
\paragraph{Individual Representation:}
A population  \(P\) consist of  \(N\) individuals, which represent quantum circuits. 
Each individual \(I_k\), \(0<k<N\), with~\(n\)~qubits and \(m\) gates per qubit is encoded as a two-dimensional list with \(n\) rows and  \(m\) columns. Note that \(n\)  and \(m\) correspond the circuit \textit{width} and \textit{depth} respectively.

Each gate \(g_j\), \(0<j<m\), acting on the qubit \(q_i\), \(0<i<n\), has following characteristics: 
\begin{itemize}
    \item \textit{name} chosen from a predefined gate set,
    \item  \textit{qubit id} to identify on which qubit the gate is applied,
    \item  \textit{affected qubits} to represent multiple-qubit gates,
    \item \textit{rotation angles} \(\theta\) for parametrized gates, e.g.~\(RX (\theta)\); if no constant value is set for the parameters, they are initialized randomly in the range~\([-\pi; \pi]\),
    \item \textit{control and target qubits} for controlled gates. 
\end{itemize}
\begin{figure}[t!]
    \includegraphics[width=0.5\textwidth]{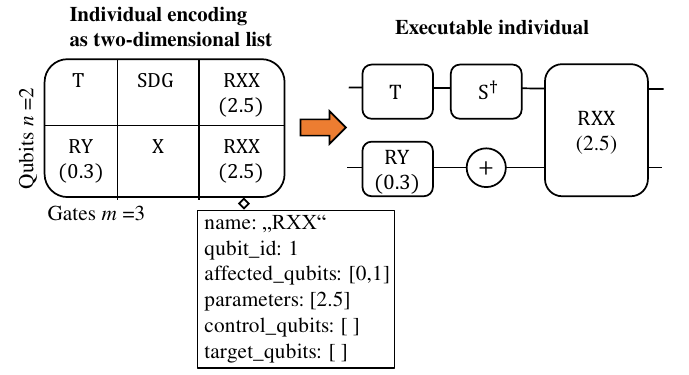}
    \caption {Individual representation: Quantum circuit on the left side with \(n=2\) qubits and \(m=3\) gates is stored as \(2 \times 3 \) python list. 
    Gates have predefined attributes, which gather metadata needed to generate the executable circuit on the right side.}
    \label{fig:Solution}
\end{figure}
\Cref{fig:Solution} depicts the described individual representation.

\paragraph{Fitness Functions:}
The fitness function defines the objective of the optimization and assigns a scalar value to each solution / individual reflecting its quality with respect to the optimization goal. A fitness function can take multiple properties of the individual into account, for example the number of gates, the quantum state it produces or its behavior in a variational training setting. It is up to the researchers to define how the different properties shall be combined and weighted to calculate the fitness. 
We've implemented several example fitness functions using our proposed framework, and we describe a selection of these next. 

\textit{Fidelity fitness function:} The aim of the fidelity fitness function is to find a quantum circuit that maximizes the fidelity to a given target state. Additional constraints can be added to the fitness function in order to find circuits matching a specific criteria, e.g. depth can be included to prefer shallow circuits, the set of quantum gates used for individual generation can be restricted, or the noise model of a specific quantum hardware can be considered.

\textit{Entanglement fitness function:} Using this fitness function, the algorithm seeks a state with maximum entanglement, i.e., a state equivalent to the Bell state for the two qubit case. Note that here the entanglement is measured with the von Neumann entropy. 

\textit{Machine learning fitness function (Classification/Regression):} In this use case, 
training accuracy on a given dataset serves as fitness. Here, we distinguish between evolution process and learning from data points, which takes place within the evolution process. As an individual learns from the dataset, parameters in parameterized gates are changed with the aim to increase training accuracy. After a certain number of learning steps, training accuracy is determined and stored as fitness value. Thus, using this fitness function, the user can discover a beneficial circuit ansatz for the given classification/regression problem. 
 
\paragraph{Selection Methods:} The framework supports three selection methods, each one can be used for the parent and survivor selection.

\textit {Random selection:} Individuals are selected randomly from the population. Here, the fitness value is not taken into account. This selection should be used when a high genetic diversity is required.

\textit {Tournament selection:} Selects $n$ individuals from the population. Therefore, the algorithm creates $n$ tournaments with a given number of individuals. In each tournament the individuals are compared with respect to their fitness. The winner of each tournament is then returned.
    
\textit {Roulette wheel selection:} Returns a population according to the roulette wheel selection approach used in genetic algorithms. The selection of an individual is a random decision, where fitness of an individual increases the probability to get selected. 

\paragraph{Crossover Methods:}
\begin{figure}[t]
%\begin{center}
    \includegraphics[width=\linewidth, height=12.2cm]{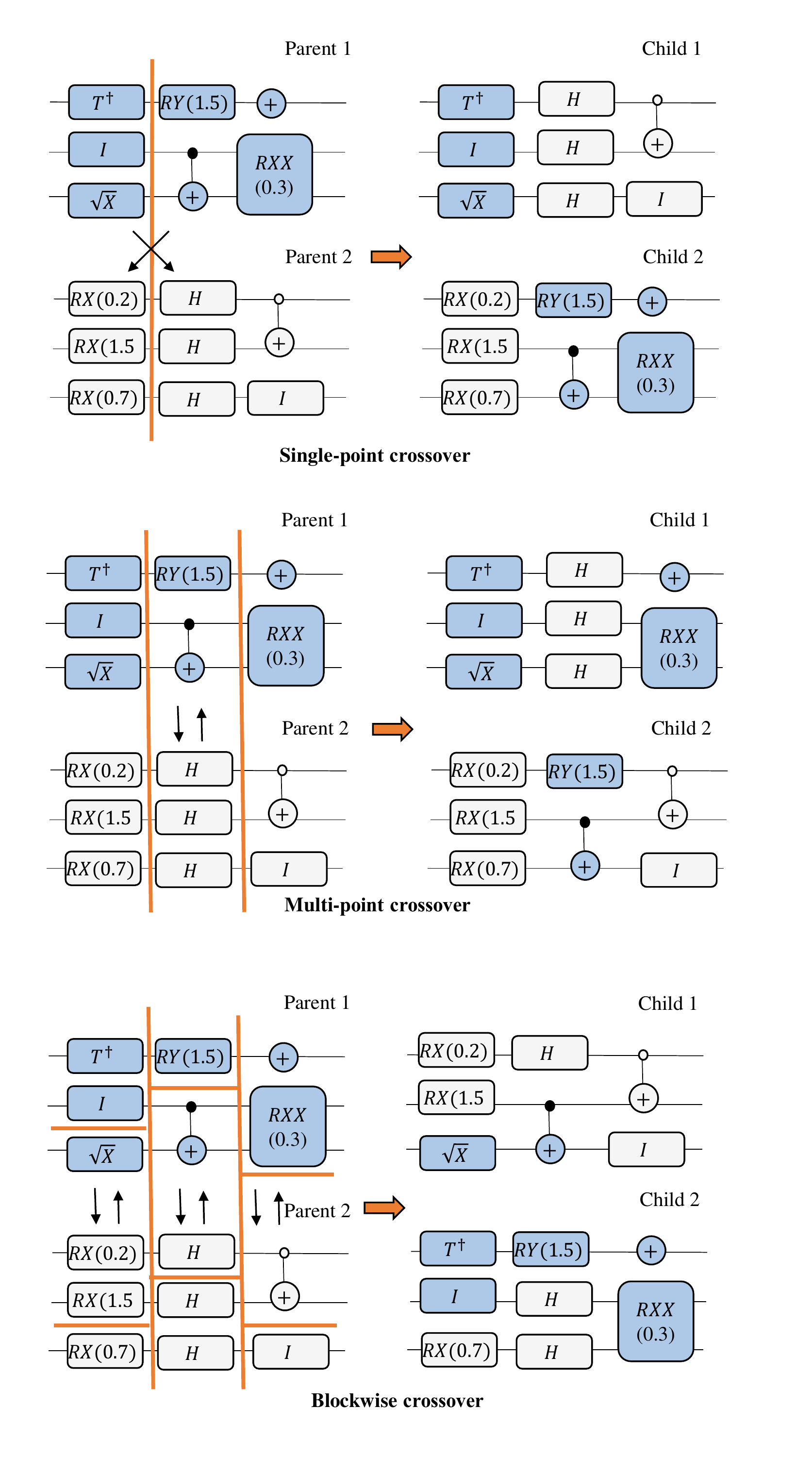}
    \caption{Crossover methods currently available in the GA4QCO framework.}
    \label{fig:Crossover}
%    \end{center}
\end{figure}
%\begin{figure*}[t!]
%\begin{center}
%    \includegraphics[width=1\textwidth]{images/Genetic_operators.pdf}
%    \caption{Crossover and mutation methods currently available in the GA4QCO framework.}
%    \label{fig:Mutations}
%    \end{center}
%\end{figure*}
We provide three crossover methods (see visualization on the Figure~\ref{fig:Crossover}), differing in complexity and resulting diversity of the genetic information.

All following crossover methods assume that the parent individuals can differ in their width and depth. Hence, the parent circuits will be padded with identity gates to meet the same size before applying the crossover logic. It is configurable whether the crossover returns all produced children or only one of them.

\textit{Single-point crossover:} Is the simplest crossover method. It chooses at the column level a random point to split each parent at and recombines the four resulting parts into two children. 

\textit{Multi-point crossover:} Differs from the first method only in the number of splitting points (more than one). Due to more randomly selected splitting points, the recombination results in more diverse mix of the parental genetic information increasing the diversity and explorative capabilities of the whole evolution process. On the other hand, this method might destroy qubit sequences, which already perform well with respect to the optimization goal.

\textit{Blockwise crossover:} Splits the parent circuits not only column-wise like the two methods before, but it chooses two-dimensional splitting points with row-wise and a column-wise dimension. While the first two crossover methods preserve all genetic information of the parents distributed across the two children, blockwise crossover will certainly destroy some multi-qubit gates due to its two-dimensional splitting. The algorithm will reconstruct broken multi-qubit gates occurring in child-solutions. That way, one parental multi-qubit gate distributed (and such broken) on the two children will result in two valid multi-qubit gates in the children after reconstruction, i.e., the all-over genetic information of the parents was changed in summa. However, this behavior is not considered negative since the evolution process works with mutating genetic information.

\paragraph{Mutation Methods:} 
 \begin{figure*}[t!]
\begin{center}
    \includegraphics[width=1\textwidth]{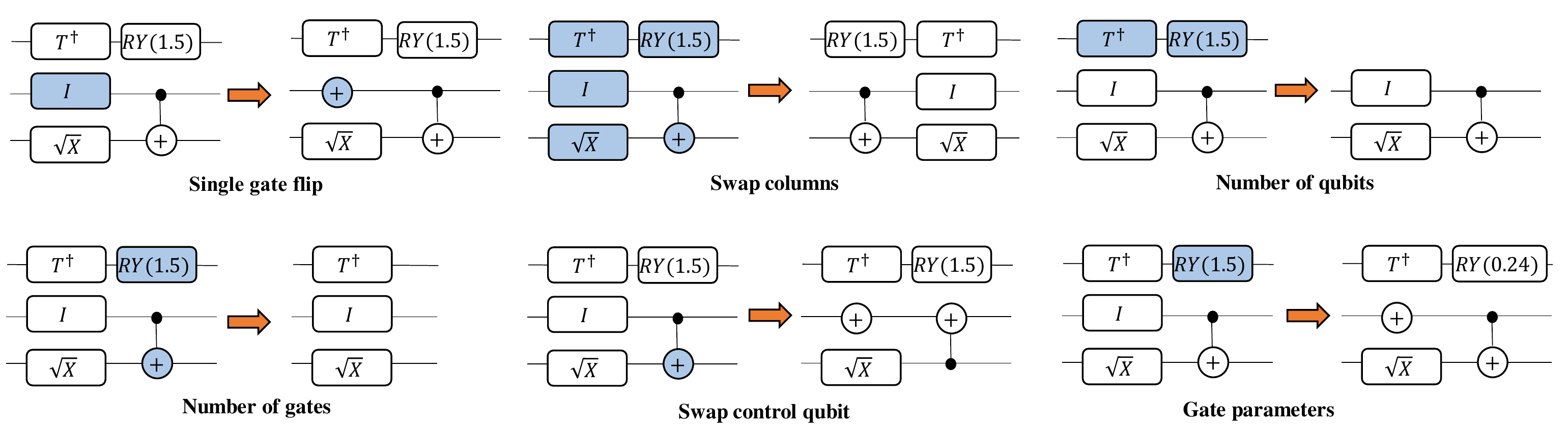}
    \caption{Mutation methods currently available in the GA4QCO framework.}
    \label{fig:Mutations}
    \end{center}
\end{figure*}
New individuals produced by a crossover operation are mutated with a specific probability.  This maintains genetic diversity and can result in children that are better than either parent wrt. the fitness function. Following mutations are provided in our framework (see Figure~\ref{fig:Mutations}): 

\textit {Single gate flip:} Randomly selects a gate and replaces it with a random gate. If the selected gate is a multi-qubit gate, all affected gates are replaced with a random gate as well.

\textit {Swap control qubit:} Randomly searches for a controlled-gate and swaps control and target qubit.

\textit {Mutate qubits} Adjusts the number of qubits randomly by adding or removing a qubit.

\textit {Mutate gates} Adjusts the number of gates randomly by adding random gates or removing gates on each qubit.

\textit {Swap columns:} Exchanges all gates from two randomly chosen columns of the circuit.

\textit {Mutate gate parameters:} Randomly selects a parameterized gate and adjusts its parameter (if such a gate is found).

\subsection{Availability}
We will be making our GA4QCO framework available to the research community on GitHub\footnote{https://github.com/PlanQK/GA4QCO} soon, including documentation and a number of tutorials. All experiments included in this paper are also available on GitHub. As described above, a core feature of the framework is its easy integration of custom fitness functions, i.e., the functionality that allows users to define their own fitness functions that will guide the evolution process to search for specific circuits. This allows the framework to be applied to a wide range of domains. We plan to adapt and extend the framework to integrate new features and fitness functions. 
\begin{figure*}[t!]
    \centering
    \includegraphics[width=1\textwidth]{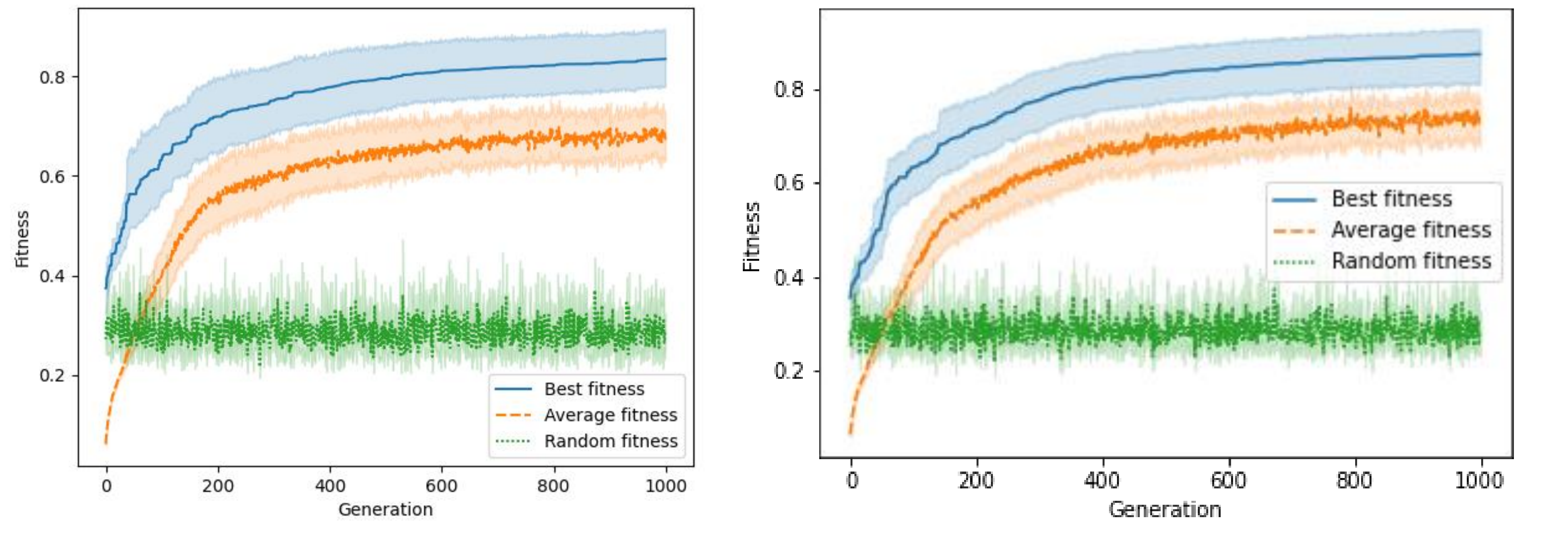}
   \caption{The best and average fitness (i.e. fidelity) compared to randomly created solutions aggregated over 10 experiments with different targets to show the mean and 95\% confidence interval. The plot on the left depicts the results without restricting the set of gates, and the plot on the right shows the results from running the same experiment with only five gates.
   }
    \label{fig:fidelity}
\end{figure*}

\section{Experiments}\label{sec:experiments}
In this section, we show results of preliminary experiments from our framework to illustrate the validity of our approach. We performed experiments using the fidelity fitness function in 2 different ways: Maximizing fidelity without any additional constraints and maximizing fidelity using a restricted gate set.
All experiments were implemented in Python 3 using Qiskit~\cite{Qiskit} and are available on the public GitHub repository, including the parameters used. For each experiment, we show the best fitness per generation, average fitness per generation as well as a random baseline. Note that the results are averaged over all experiments. The random baseline shows the best randomly created individual per generation, in each generation an equal amount of solutions is created as with the GA.

\paragraph{Fidelity:}
Recall that the aim of the fidelity fitness function is to maximize the fidelity between a given target and the state prepared by a circuit created by the GA4QCO framework. For our experiments, we used Qiskit~\cite{Qiskit} to create random circuits whose statevectors serve as the target states. 
Each circuit has 4 qubits and a depth of 20. We created 10 such circuits randomly and ran the algorithm for each with the same parameters. The number of generations was set to 1000 and a population of 200 individuals was used. A full set of parameters used is available in the GitHub repository.

The results are shown in Figure~\ref{fig:fidelity}. The fitness (i.e. fidelity) continuously improves, approaching a fidelity of 0.8. Running the algorithm longer with a larger population may even yield better results.

\paragraph{Fidelity with the restricted gate set:} 
As in the previous case, the framework searches for a circuit that prepares a quantum state close to the specified target state. 
But this time, we restrict the gates that can be used in individuals.
This experiment has a practical background, e.g. when the user knows the properties of the quantum computation device and aspire to use only the gates physically implemented on this device.

In our experiment, we restrict gates to the following set: \{'id', 'rz', 'sx', 'x', 'cx'\} and run the evolution with the same parameters as in the first example.

The results shown in Figure~\ref{fig:fidelity} demonstrate that even with the restricted gate set our framework is able to find a much better solution than a random guessing. The highest fitness of the best individuals returned after 1000 generations in 10 runs we observed was 0.994 from maximum 1.0 (in contrast to random baseline with the highest fitness 0.79), the lowest one was 0.67, which, nevertheless, is still significantly better than the average fitness of the random baseline~(0.29). 

\begin{figure*}[t!]
    \centering
    \includegraphics[width=1\textwidth]{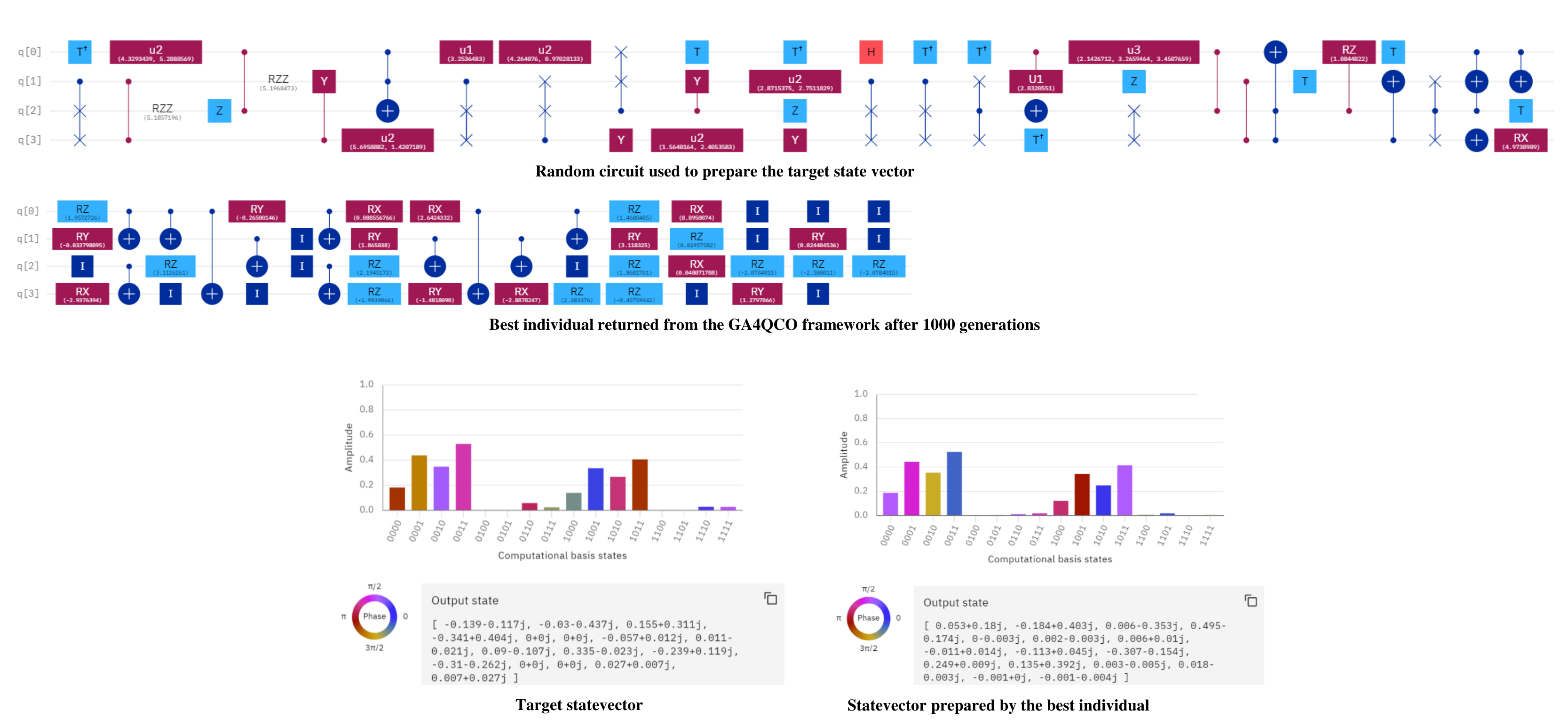}
   \caption{Results from restricted gate set experiment for a single seed. 
   Top: A randomly created circuit (seed 10) vs. the circuit of the individual with the highest fitness 0.994 (fidelity).
   Bottom: A comparison of the target-statevector and the statevector prepared by the best individual.
   }
    \label{fig:circuits}
\end{figure*}

Figure \ref{fig:circuits} demonstrates the random circuit used to prepare target statevector and the best individual as well as target and output statevectors in comparison. 

\section{Outlook}\label{sec:outlook}
The aim of the experiments from the previous section was, among other things, to show the validity of our approach, i.e., to demonstrate how GA can be applied to search for quantum circuits that satisfy some constraint or achieve some objective. These experiments could also be modified and extended. For instance, a further constraint for the fidelity fitness function could be the optimization under noise, i.e., inclusion of a noise model and noisy gates. Furthermore, the GAs used in our experiments ran for 1000 generations with a population of 200 individuals, increasing the number of generations or population size may improve the results even further. We limited the circuits to 4 qubits and a depth of 20. Investigating how the algorithm performs on larger circuits is a crucial question, especially in comparison to other approaches.

Moreover, as the framework is designed as a toolbox, we plan to extend and improve it and also integrate more advanced fitness functions.

\section{Conclusion} \label{sec:conclusion}
The design of quantum circuits is still mostly performed manually, a task which is often tedious and not intuitive. However, different approaches to automate this have been explored by the research community over the years, most notably applying genetic algorithms or reinforcement learning. In this paper, we proposed our GA4QCO framework, which uses a genetic algorithm to search, i.e., construct, the architecture of quantum circuits with specific characteristics. Our framework allows the easy integration of custom fitness functions, a feature allowing its users to apply the framework to a wide range of domains. 

We furthermore ran experiments on different problem settings to demonstrate the validity of our approach. We define target circuits and run the algorithm to search for circuits that maximize the fidelity between the constructed and target states. We also optimize the circuits using a restricted gate set. While our framework can be successfully applied to basic problems, we plan to further extend it and evaluate it on more advanced problems as there are still many open questions in the field of circuit construction.
\subsubsection*{Acknowledgments}
This work was partially funded by the BMWK project \textit{PlanQK (01MK20005F / 01MK20005I)} and Einstein Research Unit on Quantum Digital Transformation \textit{(ERU-2020-607)}.

\printbibliography

\end{document}